\documentclass[10pt]{article}

\usepackage{a4}
\usepackage[round]{natbib}
\usepackage{url}
\usepackage{charter}
\usepackage{textcomp}

\begin{document}

\title{Wikis are good for knowlegde management\footnote{This position paper was written in 2005. It has never been published apart from arXiv.org.}}
\author{Sander Spek\\elektropost@sander-s.net}
\date{Institute for Knowledge and Agent Technology\\Universiteit Maastricht}

\maketitle

\abstract{Wikis provide a new way of collaboration and knowledge sharing. Wikis are software that
allows users to work collectively on a web-based knowledge base. Wikis are characterised by a
sense of anarchism, collaboration, connectivity, organic development and self-healing, and they
rely on trust. We list several concerns about applying wikis in professional organisation. After
these concerns are met, wikis can provide a progessive, new knowledge sharing and collabora-
tion tool.}

\section{Grassroots knowledge sharing}

During the last decades, the sharing of information and knowledge has gone through an unheard
rapidisation. Both the amount of information shared, as well as the speed in which this happens,
has taken a huge flight. In many fields of progress, professional organisations were more ad-
vanced then ‘amateur knowledge sharers’. Remarkably, in some fields this has changed over the
years. On the Internet, many succesful grassroots initiatives have popped up, that still didn’t
make it to the somewhat static and controlled environment of professional business organisa-
tions. Think of free-software projects, but also of free-information projects, as wikis, weblogs,
and open news sites like Indymedia. \citep{Moller2005}

In this paper, we will examine on of the most popular and upcoming knowledge sharing
enablers on the Internet: wikis. Consequently, we will discuss why wikis are not (yet) adequate
for a professional environment, and what could be changed to have companies adapt the power
of wiki-based knowledge sharing too.

\section{Wikis}

\citet{Moller2005}[page vii, translated] defines wikis as “open websites that can be edited by every
visitor.”\footnote{The original definition is: “offene Websites, die jeder Besucher bearbeiten kann.”} We will extend this definition in two ways: (1) we explicitly state that modifying can
also include deleting information, and (2) we mention the fact that a wiki is in fact a knowledge
base. Hence, we come up with the following defition of a wiki:

\begin{quotation}

\noindent a piece of software that allows users to add, modify, and/or delete information from a
knowledge base via the web

\end{quotation} \normalfont
  
The exact functionality of the wiki differs per software package~\--- the
so-called wiki engine~\--- and the configuration of the engine. For instance,
some wikis allow every anonymous visitor to alter the information, where as
others only allow registered users to do this. Some engines allow multimedia
uploads (like images or sound files), where others have functionality to
prevent `edit conflicts'.

Wikis are invented by Ward Cunningham, and elaborated upon by \citet{Leuf01}.

\subsection{Characteristics}

\label{characteristics}

We can distinguish six prominent characteristics of wikis. Wikis are by nature
(1) anarchistic, (2) collaborative, (3) connected, (4) organic, (5)
self-healing, (6) based on trust.

\subsubsection*{Anarchistic}

Wikis are anarchistic in the sense that there is no power structure. In
general, no user has more rights then any other user.  On many wikis, anonymous
users have the same rights as registered users. Sometimes some power structure
is established. For instance, on Wikipedia there are sysops (system operators)
that have additional functionality for the revertion of vandalism. Because of
the anarchistic nature, a power structure can lead to conflicts between users,
e.g., when assigning new sysops.

Because of the equality of rights, there is also no division of labour. There
is no director that tells subordinates what to do.  Each individual can select
the role that best fits his or her preferences.

\subsubsection*{Collaborative}

A wiki is for the most part a collaboration tool. The users work together to
create a certain end-product, whether that be a report, a reference guide, or
an encyclopedia. Its strength is in its ability to facilitate users to
cooperate without a division of labour has been made in advance. With support
for so-called `edit conflicts' users can even work on the same page on the same
time.

A wiki is less good at supporting communication between users. Many wikis also
contain discussion pages, but for elaborate discussions other tools, like
e-mail or web-based forums, provide better functionality.

\subsubsection*{Connected}

The pages in a wiki generally are not ordered in a prescribed way, like in a book. Just like on
the World Wide Web, wikipages are interlinked in a network structure. Organising knowledge
in a network instead of a sequence appeares to be a natural way of knowledge representation, as
also knowledge in books on its turn can be seen as networked knowledge. Or as \citet{Focault1972}[p. 23] puts it: “The frontiers of a book are never clear-cut: beyond the title, the first lines, and
the last full stop, beyond its internal configuration and its autonomous form, it is caught up in a
system of references to other books, other texts, other sentences: it is a node within a network.”

Creating links in a wiki is generally very easy: applying the linking syntax to a certain term,
usually putting it between brackets or writing it in camel case\footnote{The early wikis used camel case for linking. Camel case is glueing words together and replacing spaces by a capital-
ization of the next character. Examples of this are camelCase (lower camel case, starting with a lower case character) and WikiWikiWeb (upper camel case, starting with a capital). Text and titles with camel case terms might look odd to the average reader, and some claim the success of modern wikis is partly because of the abandoning of camel case.}, immediately creates a link to a
wikipage which has that term as a title.

\subsubsection*{Organic}

Because of the lack of control and delegated division of labour, a wiki expands
itself in an organic way. Information on one subject can be very detailled,
whereas other equally important subjects might get not elaborated upon. The
direction in which a wiki expands depends largely on its community, with all
its \--- perhaps uncommon \--- interests, hypes and trends. For instance, the
Dutch Wikipedia has an very active user living in Thailand.  For this reason,
the information on Thailand in the Dutch Wikipedia is unproportianlly large,
compared to, e.g., the African countries.

\subsubsection*{Self-healing}

Wikis are sensitive to vandalism: malicious users deleting information or
inserting incorrect information. Especially wikis that allow unregistred users
to make changes encounter this problem. However, a wiki, especially wikis with
a large community, have a high potential of self-healing. The famous citation
by \citet{Raymond97} on open-source software development, ``Given enough eyes,
all bugs are shallow'',\footnote{Raymond calls this ‘Linus’s Law’.} applies to wikis too.  As a lot of people read the
pages, if the community is large enough amongst them might be experts in the
fields, vandalism and incorrect information are likely to be corrected in a
fair amount of time.

Generally in the large Wikipedias, reverting obvious vandalisms is a matter of
seconds to minutes. The real problem however lies with incorrect information
that looks plausible or is generally believed to be plausible. Wikis might be
sensitive to hoaxes and urban myths. However, the German Wikipedia shows to
outperform the more established Encarta and Brockhaus, even concerning content
quality \citep{Kurzidem04}.

\subsubsection*{Based on trust}

Wikicommunities heavily on the trust between its users. Since it is so easy to
revert changes, conflicts can quickly result in `edit wars' where multiple
users keep on reverting each others changes because they don't agree. And
because of the anarchistic character, there is no arbitrator to make a final
decision. Of course they community can create a consensus, e.g., by voting, but
there is no way to enforce the result of the consensus upon the users in
conflict.

\subsection{Advantages}

From the characteristics in the previous subsection, we can tell that wikis have several advan-
tages over other collaboration tools. Wikis allow users to work on the same document, or the
same part of the document, at the same time. And users can at any time consult the latest version
of the document, without the editors having to send out copies of the new version.

Another important advantages is that it takes away the role of editor, or rather: it makes
everyone an editor. It is quicker and easier to do then any other publish method, and hence it
expands quicker and is more up to date. And in most cases, its self-healing capabilities function
well enough to remove or correct the ‘junk’ that gets uploaded.

Each user can, within certain limitations, select his/her own role. This means that users
get more satisfaction out of their work and can be more motivated. Also, giving users certain
freedom stimulates the creative process, which can result in unexpected new findings. Hypothet-
ically, this creative process can even transcend organisational boundaries, so that people from
different departments can work on documents or concepts that turn out to be relevant to both
departments.

\subsection{Successful examples}

\label{sec:examples}

On the Internet, there exist several successful wikis. The most well-known and
undoubtly most successful example is Wikipedia, a descendant of the Nupedia
project.\footnote{Nupedia was a project to create an open-content encyclopedia
with a peer-review system. Its main editor, Larry Sanger, resigned in 2002, and
in 2003 the website went down.} Wikipedia is an effort to create an open
encyclopedia in several languages by means of wiki software, started by Jimmy
Wales. The content of the encyclopedia is available to users under the GNU Free
Documentation License. The English language Wikipedia \--- with over 440.000
articles\footnote{Last checked on the 4th of January, 2005.} \--- is larger
than the Encyclop\ae dia Brittanica \--- with 120.000 articles in the on-line
version, the biggest edition.\footnote{According to
\url{http://en.wikipedia.org/wiki/Encyclopedia_Brittanica}.}

Wikipedia uses the MediaWiki software. There are several Wikipedia-related
projects that also use this wiki engine, such as Wiktionary (dictionary),
Wiki\-Books (textbooks and manuals), Wiki\-Quote,
Wiki\-Source (previously published documents) and Wiki\-News.

Other well-known wikis include are the Meat\-Ball\-Wiki (about on-line culture
and communities), the Linux\-Wiki, Wiki\-Travel (a travel guide), and the
Switch\-Wiki, which aims to be a list of all available wikis around the globe.

\section{Applying Wikis in Organisations}

All successful examples of wiki implementations mentioned in section
\ref{sec:examples} are freely available on the Internet, and its user community
consists completely of volunteers. Wikis are now gaining
attention in professional organisation, and companies like Socialtext and JotSpot now provide
wiki services to companies (see section \ref{sec:overviewprof}). The application of wikis in business might pro-
vide a new way of knowledge sharing and might connect people with similar interest that are
organisationally dispersed. However, before implementing the software straight away in a busi-
ness environment, we see a few points of attention. We will discuss them in four groups: (1)
motivational considerations, (2) authoritan considerations, (3) strategic considerations, and (4)
effectivity considerations

\subsection{Motivational concerns}

The volunteers that fill the wikis on the Internet do not gain much by doing
so. The users might add their knowledge because they expect to get other
knowledge in return.\footnote{Please note that this section is higly
speculative. Without proper research one cannot claim anything about the
motives of strangers. Pitfalls, like `projection'~\--- the ascription of the
authors own motives to the members of the wiki community~\---, are luring.} The
volunteers that edit and add knowledge, often also use the wiki for their own
reference. However, when a wiki grows large enough, we can see a `prisoner's
dilemma'\footnote{A prisoner's dilemma is a situation in which the best
rational decision for single individuals turns out to result in a suboptimal
result for the group. This best rational strategy can be a so-called `dominant
strategy', a strategy that is best whatever the other player does.  \citep[p.
76\--77]{DoumaSchreuder98}} developping: one user's contributions do not make a
significant difference to the size or quality of the wiki.  Therefore, the
benefits are available anyhow, whether the user makes an effort to add more
knowledge or not.\footnote{In smaller wikis this symptom might not appear,
because a user feels he or she makes a significant contribution to the wiki.
And when a wiki grows, more users might feel attracted to it and start to share
too. Thus, on a small wiki a user might feel the contributions are beneficial
for gaining own knowledge.} This situation resembles well-known economic
examples of prisoner's dilemmas, e.g., an inhabitant not donating voluntary
money to its country's security system, because the total outcome of the money
collection will not be significantly affected by the individual's decision
whether or not to donate, and therefor the individual's expected benefits are
the same in any case.

An elucidation of the contributor's motivation of large wikis should therefore
be sought in other area's. Users can feel good sharing knowledge that other
people might need~\--- a philantrophic motivation~\---, they might enjoy taking
credit for their well-written articles, or they might enjoy being a member of a
certain community. Some Wikipedia contributors maintain a list of articles
they've worked on, and frequently well-written articles are honoured, e.g., by
a writing contest or by announcing them on the main page of the wiki. These are
clues that might indicate the second option. Indications of the third options
are the lively communities existing on many wiki, including non-wiki-related
chats\footnote{Some wikis even have a designated area for off-topic chats, like
the \emph{Village pump} on the English Wikipedia, and \emph{De kroeg} (the pub)
on the Dutch Wikipedia.}, and real-life meetings.

In a professional environment, these motivations might change. A career holds a
high priority to people in most modern societies, and many professionals
maintain a competitive attitude towards co-workers. Knowledge is power, and
this yields no place for philantrophic attitutes. An encyclopedia as Wikipedia
contains mainly basic knowledge in expert fields, and general practicioners and
legal experts that in their free time share this basic knowledge on Wikipedia
will not loose much power. In a professional organisation, the knowledge to be
shared is more focussed and therefore more inclined to go in-depth, resulting
in a higher power-loss risk.

Professional organisations that want their knowledge base (their wiki) filled
with the knowledge that lives within the organisation have two main options.
First, they can stimulate or reinforce their employees to add knowledge to the
wiki. This can, for instance, be done by setting periodic quota that employees
have to meet, or to reward qualitative or quantitative contributions made by
employees. These measures tend to stimulate quantity over quality, so
management should well think through the implications of the measures the
enforce. The second option is the disolvement of the prisoner's dilemma as much
as possible. This can, for instance, be done by clustering a large wiki, or by
setting special fields of attention. In this way, the user becomes more aware
of certain interesting areas in the wiki, by which useful contributions from
co-workers become more visible. Also, this might improve the community-feeling
in certain parts of the wiki.

\subsection{Authoritan concerns}

Organisations are generally build around a certain authoritan model, where
certain people (usually managers) have responsibility for subparts of the
organisation, or the organisation as a whole in the case of top management, and
delegate tasks to subordinates. During the years the models of organisations
have changed, going from hierarchical pyramids via networked organisation with
high employee autonomy back to a sort of hierarchical diamond. However, the
concepts of resposibility and delegating tasks have always been in place.

As discussed in subsection \ref{characteristics}, wikis are anarchistic by
nature. In a pure wiki, there are no users with a higher authority as others,
and each individual picks its own tasks.\footnote{However, there might be peer
pressure to perform additional tasks. For instance, contributors can be
encouraged by fellow contributors to pay more attention to their spelling
errors, or to add keywords for classification to their new articles. However,
these requests have to be made in a friendly and co-operative way since there
are no methods of enforcement.} This might conflict with the labour divisions
that already have been established within an organisation. Managers are not
likely to give up their possibilities of delegating tasks, and moreover, in
discussions regard content people might use their position in the organisation
as an argument, thereby overruling `real' arguments. 

\subsection{Strategic concerns}

Organisations usually like to have control of the direction in which they are
developing. For this reason, it creates a mission statement, a strategy, and
goals. \citep[p.40\--44]{Daft98} A wiki, on the contrary, develops in an
organic way. Since there generally is no authority, developments cannot be
affected in a top-down fashion. However, sometimes bottom-up developments can
steer the information expansion, e.g., by organising a `theme week' in which
contributors are stimulated to focus their attention on a (yet undervalued)
topic. For wikis to become useful in organisations, more control over the
knowledge expansion is needed.

\subsection{Effectivity concerns}

One concern of large organisation is division in departments and units. This
division is needed to keep the organisation managable, but at the same time it
creates barriers between people that might work in related areas, and the
organisation would benefit from knowledge sharing between those people. The
trend of organisations adopting `people finder' applications, that allow
employees to search for other people based on their areas of expertise, reflect
this need. Wikis might also break down these barriers. When several people of
different departments work on a specific set of pages related to their area of
work, they will get to know each other and each other's expertises. However, it
is still a question whether a wiki can break down these barriers, or that the
organisational boundaries will also create matching clusters of pages without
much connections between the clusters.

\section{Overview of wikis in an professional environment}

\label{sec:overviewprof}

Wikis are getting more and more noticed by the professional world, as we can tell from the
appearance of management articles about wikis, like \citet{Rand2004} and \citet{Sharma2004}, and the
note about wikis in Gartner’s ‘Hype cycle for emerging technologies’ \citep{Bradshaw2004}.
Several companies offer commercial services that, to some extend, resemble wikis. Social-
Text\footnote{\url{http://www.socialtext.com}} offers a software package which provides a simple wiki interface. Users can post notes, in a log style, and integrate this with their e-mail. It is aimed at unstructured, ad hoc collaboration, and can therefore be positioned somewhere between wikis and collaborative weblogs. In
general, wikis are more stuctured and persistend then SocialText. A competing company called
JobSpot\footnote{\url{http://www.jobspot.com}} provides an easy interface, with WYSIWYG editor, to access a relational database that
keeps contact information of clients and maintains a calendar. Other database applications can
be created. JobSpot focusses more on storing data, than on creating documents. Netomat\footnote{\url{http://www.netomat.net}} offers
multimedia whiteboards for real-time collaboration. Users can collaborate using many types of
multimedia, but the knowledge isn’t stored in a manner that allows retrieval at a later point. All
three commercial products have some flavour of wikis, but are not exactly it.

On the open-source side of wiki developments, a wiki engine called TWiki\footnote{\url{http://www.twiki.org}}
is geared more
towards a professional application then other wiki engines. For instance, it allows the creation
of forms so that users can easily enter data that will be grouped on wiki pages. Also, the best
known wiki engine, MediaWiki, is used by several companies, like Gartner and Novell.\footnote{According to \url{http://meta.wikimedia.org/wiki/Sites_using_MediaWiki}.}

\section{Conclusions}

In this paper, we took a look at a new, original, and increasinly popular piece of collaboration
software: wikis. They have a six prominent characteristics, namely anarchistic, collaborative,
connected, organic, self-healing, and based on trust. Wikis provide new ways of working and
can turn out to be beneficial to professional organisations. Several commercial services that
have a wiki flavour already exist.

However, for wikis to become succesful in professional organisations and to provide an
added value in the organisation’s knowledge management, several modifications concerning the
software and its surrounding policy have to be considered. We have elaborated upon motiva-
tional, authoritan, strategic, and effectivity concerns. For a wiki to be succesfull in professional
organisations, these concerns have to be met by changes in the software or by attention and
action of the management

\bibliographystyle{natbib}
\bibliography{literature}

\end{document}